\documentclass[prb,twocolumn,superscriptaddress,preprintnumbers]{revtex4}

\usepackage{graphicx}
\usepackage{dcolumn}
\usepackage{amsmath}
\usepackage{color}

\begin{document}

\title{Electronic correlations and long-range magnetic ordering in NiO tuned by pressure}

\author{G. M. Gaifutdinov}
\affiliation{M. N. Mikheev Institute of Metal Physics, Russian Academy of Sciences, 620108 Yekaterinburg, Russia}
\affiliation{Ural Federal University, 620002 Yekaterinburg, Russia}

\author{I. V. Leonov}
\affiliation{M. N. Mikheev Institute of Metal Physics, Russian Academy of Sciences, 620108 Yekaterinburg, Russia}
\affiliation{Ural Federal University, 620002 Yekaterinburg, Russia}
\affiliation{Skolkovo Institute of Science and Technology (Skoltech), 121205, Moscow, Russia}

\begin{abstract}

Using the DFT+dynamical mean-field theory method we revisit the pressure-temperature phase diagram of the prototypical correlated insulator NiO. We study the pressure-induced evolution of the 
electronic structure, magnetic state, and exchange couplings of the antiferromagnetic phase of NiO. We calculate the ordered magnetic moments and uniform spin susceptibility of the Ni $3d$ states of 
NiO, which allow us to determine the pressure-dependence of the N\'eel temperature $T_N$.
We note that the long-range magnetism has no significant effects on the valence band photoemission spectra of NiO under moderate compressions, 
implying the importance of correlations effects to explain the insulating state of NiO.
The antiferromagnetic insulating state is found to be stable up to the high compression value $\sim$0.4~$V_0$ (assuming the cubic $B1$ crystal structure of NiO), and is associated with a 
(correlated-assisted) Slater insulating state driven by the long-range magnetic ordering. In fact, the paramagnetic phase of NiO at such high compression is found to be metallic, implying 
delocalization of the Ni $3d$ states.
The calculated $T_N$ exhibits a non-monotonic behavior upon compression, with a maximum associated with the crossover from Mott localized (strong coupling) to itinerant moment regimes, in 
qualitative agreement with the phase diagram of the half-filled single-band Hubbard model.
We point out the importance of the non-local correlation effects to explain the magnetic properties of NiO. 

\end{abstract}

\maketitle
\nobreakspace

\section{Introduction}

In the last several decades much attention has been devoted to understanding of the electronic structure and magnetic properties of materials with strongly correlated electrons 
\cite{Mott_1990,Imada_1998,Anisimov_2010,Khomskii_2014}. In such systems, the complex interplay between electronic correlations and the spin, orbital, and lattice degrees of freedom leads to a 
wealth of ordering phenomena and emergent phases, making them promising for fundamental research and technological applications. A particular example of such emergent behavior is the Mott 
insulator-to-metal phase transition in a series of transition metal oxides, e.g., in MnO, FeO, CoO, and NiO, driven by high pressure. Over the past decades these materials have served as the 
prototype materials with strongly correlated electrons that exhibit the Mott-Hubbard or charge-transfer insulating behavior, associated with strong localization of the $3d$ electrons 
\cite{Mott_1990,Imada_1998,Anisimov_2010,Khomskii_2014,Zaanen_1985}. It has been a long-standing challenge in condensed matter physics to understand theoretically the interplay between electronic 
correlations, magnetism, and lattice (e.g., to explain structural polymorphism) in these materials.

Nickel monoxide NiO is among the archetypical materials which play a key role in the investigations of the electronic and magnetic properties of strongly correlated electron systems 
\cite{Mott_1990,Imada_1998,Anisimov_2010,Khomskii_2014}. It is a correlated insulator with a large energy gap of $\sim$4 eV as determined from photemission \cite{Sawatzky_1984,Shen_1991}, whereas 
the optical absorption measurements show an onset of absorbtion at 3.1 eV with a maximum at 4.3 eV \cite{Powell_1970}. Below the N\'eel temperature $T_N \sim 523$~K NiO exhibits a long-range 
antiferromagnetic (AFM) ordering (type-II) of Ni ions with the local magnetic moment of 1.77 $\mu_B$ (at ambient conditions) \cite{Roth_1958,Hutchings_1972,Shanker_1973}. It has a 
face-centered-cubic crystal structure (NaCl-type, space group $Fm\bar{3}m$) in the paramagnetic (PM) phase. Below $T_N$ a structural phase transition takes place to the distorted rhombohedral 
structure (space group $R\bar{3}m$), characterized by a small contraction along the trigonal $(111)$ axis of the cubic structure \cite{Rooksby_1948,Roth_1958,Eto_2000}. 

Over the past decades much attention has been devoted to understanding the nature of the band gap and low-energy excitations spectra in NiO \cite{Sawatzky_1984,Shen_1991,Shukla_2003,Schuler_2005}. 
Experiments suggest that NiO is a charge-transfer insulator in which the top of the valence band is primarily formed with the O $2p$ states, while the bottom of the conduction band is of the Ni $3d$ 
character \cite{Shukla_2003,Schuler_2005}. The insulating state is characterized by a large band gap of about 4 eV, associated with strong on-site Coulomb repulsion between the Ni $3d$ electrons 
\cite{Sawatzky_1984,Shen_1991}. Due to the highly correlated behavior of the Ni $3d$ states, theoretical computations of the electronic structure and magnetic properties of NiO (as well as other 
strongly correlated materials) have proven to be particularly difficult \cite{Anisimov_1991,Anisimov_1993,Anisimov_1997a,Dudarev_1998,Feng_2004,
Cococcioni_2005,Zhang_2006,Fischer_2009,Schon_2012,Trimarchi_2018,
Bredow_2000,Liu_2020,Aryasetiawan_1995,Li_2005,Kotani_2007,Kotani_2008,
Rodl_2009,Karlsson_2010,Jiang_2010,Rodl_2012,Sakuma_2013,Reining_2018,
Pokhilko_2023,Ovchinnikov_2021,Orlov_2023}. 
It turns out that for a reliable description of the electronic properties of NiO it is necessary to take into account the effects of dynamical correlations and material specific details of the band 
structure of NiO (e.g., a momentum dependence of the electronic states, crystal-field splitting of the $3d$ bands, and charge transfer $p$-$d$ gap) 
\cite{Kotliar_2004,Georges_1996,Kotliar_2006,Held_2007}.

In our study, we employ the DFT+dynamical mean-field theory (DFT+DMFT) method \cite{Kotliar_2004,Georges_1996,Kotliar_2006,Held_2007} implemented with full self-consistency over the charge density 
\cite{Pourovskii_2007,Haule_2007,Aichhorn_2009,DiMarco_2009,Amadon_2012,Park_2014,Hampel_2020,Leonov_2015a,Leonov_2015b,Leonov_2016,Leonov_2020,Koemets_2021} to compute the electronic structure of 
strongly correlated electron materials. It has been shown that applications of DFT+DMFT capture all generic aspects of the effects of (local) dynamical correlations near the Mott insulator-to-metal 
phase transition (IMT) such as a coherent quasiparticle behavior, large spectral weight transfer near the Fermi level, and strong renormalizations of the effective electron mass 
\cite{Kotliar_2004,Georges_1996,Kotliar_2006,Held_2007}. The effects of (local) dynamical correlations are quantified by self-energy, which is determined self-consistently within DFT+DMFT on the 
Matsubara frequency domain \cite{Note1}. Using DFT+DMFT it becomes possible to compute the material-specific properties of complex correlated materials, e.g., to determine the electronic structure, 
magnetic state, and crystal structure of materials at finite temperatures, e.g., near the Mott transition \cite{Kotliar_2004,Georges_1996,Kotliar_2006,Held_2007}. Indeed, previous DFT+DMFT 
calculations for the PM phase of NiO \cite{Wan_2006,Kunes_2007a,Kunes_2007b,Yin_2008,Karolak_2010,Zhao_2012,Nekrasov_2012,Nekrasov_2013,Leonov_2016,
Panda_2016,Hariki_2017,Luder_2017,Lechermann_2019,Mandal_2019,Lanata_2019,Ghiasi_2019,Leonov_2020,Hariki_2020,
Leonov_2021} successfully explain the correlated charge-transfer insulating state accompanied by strong localization of the Ni $3d$ orbitals, in agreement with experiments 
\cite{Sawatzky_1984,Shen_1991,Shukla_2003,Schuler_2005}. 

However, under high pressure conditions the experimental and theoretical situation is less clear. Using the x-ray diffraction and transport measurements it was shown that the IMT in NiO takes place 
upon compression above $\sim$240 GPa\cite{Gavriliuk_2012,Gavriliuk_2023}. On the other hand, combined neutron forward scattering and x-ray diffraction experiments confirm the antiferromagnetic state 
of NiO to persist up to 280 GPa, the highest pressure where magnetism has been observed so far, ruling out the collapse of local magnetic moments (delocalization of the Ni $3d$ states) in NiO 
\cite{Potapkin_2016}. 
In qualitative agreement with this, previous DFT+DMFT calculations suggest that PM NiO exhibits the Mott IMT under high compression $\sim $0.54~$V_0$ (above 429 GPa) \cite{Leonov_2016,Leonov_2020}. 
The phase transition is accompanied by delocalization of the Ni $3d$ states (collapse of local magnetic moments). It results in a discontinuous change of the lattice volume by $\sim$1.4\%, implying 
a complex interplay between chemical bonding and electronic correlations. In the same work \cite{Leonov_2016,Leonov_2020}, the pressure-driven Mott IMT and magnetic collapse phenomena were 
successfully explained in the PM state of the series of MnO, FeO, and CoO, in agreement with available experiments. 
In these DFT+DMFT calculations the on-site Hubbard $U=10$ eV and Hund's exchange $J=1$ eV parameters were set for NiO to fit the energy gap value at ambient conditions, and were assumed to be 
pressure-independent \cite{Leonov_2016,Leonov_2020}. The DFT+DMFT calculations with the sufficiently smaller Hubbard $U = 8$ eV ($J = 1$ eV) gave a reduced transition pressure $\sim$248 GPa, 
seemingly consistent with Refs.~\cite{Gavriliuk_2012,Gavriliuk_2023}. Note however that these calculations (with $U = 8$ and $J = 1$ eV) yield a relatively small gap value of $\sim$2 eV at ambient 
pressure, contradicting to experiment \cite{Sawatzky_1984,Shen_1991}. 

In our paper, we revisit the pressure-temperature phase diagram of the prototypical strongly correlated antiferromagnetic insulating material, NiO, using the spin-polarized DFT+DMFT method 
\cite{Kotliar_2004,Georges_1996,Kotliar_2006,Held_2007}. In particular, we study the effects of electronic correlations and long-range magnetic ordering on the electronic structure, magnetic state, 
and exchange couplings of NiO under pressure. Using DFT+DMFT we compute the pressure and temperature dependence of the uniform static spin susceptibility and ordered magnetic moments of Ni ions and 
determine the pressure dependence of the N\'eel temperature of NiO. Our results show the importance of correlation effects to determine the electronic structure and magnetic properties of NiO upon 
high compression.

\section{Method}

In this work, we study the electronic structure and magnetic properties of the long-range magnetically ordered $B1$ phase of NiO under pressure using the DFT+DMFT computational approach 
\cite{Kotliar_2004,Georges_1996,Kotliar_2006,Held_2007}. We employ a fully charge self-consistent DFT+DMFT approach 
\cite{Pourovskii_2007,Haule_2007,Aichhorn_2009,DiMarco_2009,Amadon_2012,Park_2014,Hampel_2020} implemented with plane-wave pseudopotentials 
\cite{Leonov_2015a,Leonov_2015b,Leonov_2016,Leonov_2020,Koemets_2021}. In DFT we use generalized gradient approximation with the Perdew-Burke-Ernzerhof exchange-correlation functional (GGA) 
\cite{Perdew_1996}  and ultrasoft pseudopotentials as implemented in the Quantum ESPRESSO package \cite{Vanderbilt_1990,Baroni_2001, Giannozzi_2009,Giannozzi_2017}. In our DFT+DMFT calculations we 
explicitly include the partially filled Ni $3d$ and O $2p$ states by constructing a basis set of atomic-centered Wannier functions within the energy window spanned by these bands 
\cite{Marzari_2012,Anisimov_2005,Trimarchi_2008,Korotin_2008}. This allows us to take into account a charge transfer between the partially occupied Ni $3d$ and O $2p$ states, accompanied by strong 
correlations of the Ni $3d$ electrons. 
In order to treat the effects of electron correlations in the partially occupied Ni $3d$ shell the low-energy (Kohn-Sham) Hamiltonian $\hat{H}^\mathrm{KS}_{\sigma, mm'}(\bf k)$ constructed in the Ni 
$3d$ and O $2p$ Wannier basis set is supplemented with the on-site Hubbard interaction term for the Ni $3d$ orbitals (in the density-density approximation):
\begin{eqnarray}
\label{eq:hamilt}
\hat{H} = \sum_{\bf{k},\sigma} \hat{H}^{\mathrm{KS}}_{\sigma,mm'}({\bf{k}}) + \frac{1}{2} \sum_{\sigma\sigma',mm'} U_{mm'}^{\sigma\sigma'} \hat{n}_{m\sigma} \hat{n}_{m'\sigma'} - 
\hat{V}_{\mathrm{DC}}.
\end{eqnarray}
Here, $\hat{n}_{m\sigma}$ is the occupation number operator (diagonal in the local basis set) with spin $\sigma$ and orbital indices $m$. $U_{mm'}^{\sigma\sigma'}$ is the reduced density-density 
form of the four-index Coulomb interaction matrix: $U_{mm'}^{\sigma\overline{\sigma}}=U_{mm'mm'}$ and $U_{mm'}^{\sigma\sigma}=U_{mm'mm'}-U_{mm'm'm}$. The latter are expressed in terms of the Slater 
integrals $F^0$, $F^2$, and $F^4$. For the $3d$ electrons these parameters are related to the on-site Hubbard interaction and Hund's exchange as $U=F^0$, $J=(F^2+F^4)/14$, and $F^2/F^4=0.625$. 
$\hat{V}_\mathrm{DC}$ is the double-counting correction to account for the electronic interactions described within DFT.

In agreement with previous studies, we use the on-site Hubbard $U = 10$ eV and Hund's exchange $J = 1$ eV for the Ni $3d$ orbitals \cite{Leonov_2016,Leonov_2020}. In our calculations we neglect the 
spin-orbit coupling which is expected to be small for the half-filled Ni $e_g$ bands. In DFT+DMFT, the quantum impurity problem is solved using the continuous-time quantum Monte Carlo (segment) 
method (CT-QMC) \cite{Werner_2006,Gull_2011}. In CT-QMC we use global updates, exchanging the spin up and down configurations for the Ni $3d$ orbitals to improve convergence of DFT+DMFT for a 
long-range magnetically ordered state (this drastically improves ergodicity of stochastic sampling within CT-QMC). In DFT+DMFT we utilize the fully localized (non-spin-polarized) double-counting 
correction evaluated from the self-consistently determined local occupations. In order to compute the {\bf k}-resolved spectra we perform analytic continuation of the self-energy results using 
Pad\'e approximants.

We note that different implementations of the DFT+DMFT method exist to treat electronic correlations in materials with a long-range magnetic ordering. In our calculations, we use DFT+DMFT 
implemented with the non-magnetic DFT band structure approach (with the non-spin-polarized exchange correlation functional in DFT: $\hat{H}^\mathrm{KS}_{\sigma, mm'}$ in Eq.~\ref{eq:hamilt} is 
non-spin-polarized). In this case, a long-range magnetic ordering sets in due to the spin-polarized self-energy contribution (due to the Hubbard term in Eq.~\ref{eq:hamilt}), evaluated 
self-consistently within DMFT. Using this DFT+DMFT approach it becomes possible to determine the N\'eel temperature of NiO. We also use the DFT+DMFT approach implemented with exchange splitting of 
the valence band structure in the spin-polarized DFT, known as the LSDA+DMFT method \cite{Minar_2011,Park_2015,Sanchez-Barriga_2017,Chatterjee_2021,DiMarco_2009} (in our case we employed the 
spin-polarized GGA calculations within DFT). However, this approach (with the same parameters $U=10$ eV and $J=1$ eV) yields unrealistically high values of the transition temperature, $T_N$. 
In addition, we notice that within LSDA+DMFT a significantly smaller value of the Hubbard $U$ is required to obtain the same energy gap in the AFM state as that in the PM phase. It seems unrealistic 
to describe the spectral properties of NiO, e.g., the energy gap value, using the same set of the Hubbard $U$ and Hund's exchange $J$ parameters for the PM and the AFM states. One thus faces a 
dilemma to reconcile that the microscopic parameters $U$ and $J$ depend on a magnetic state.

In the paramagnetic phase NiO has a cubic $B1$ (rocksalt) crystal structure (space group $Fm\bar{3}m$). NiO is antiferromagnetic (AFM type-II) below the N\'eel temperature of $\sim$523 K 
\cite{Roth_1958,Hutchings_1972,Shanker_1973}. It is well established experimentally that AFM NiO has a distorted rhombohedral structure (space group $R\bar{3}m$), characterized by a small 
contraction along the trigonal $(111)$ axis of the cubic structure \cite{Rooksby_1948,Roth_1958,Eto_2000}. In our study we neglect this small distortion and consider the rhombohedral supercell of 
the cubic $B1$ crystal structure, with two formula units of NiO. The DFT+DMFT calculations are performed for NiO with a long-range magnetic ordering (AFM type-II). We explore the effects of lattice 
compression and finite temperatures on the electronic structure and magnetic properties of NiO. 
The compressed phase is denoted by the relative volume with respect to the calculated equilibrium lattice volume evaluated previously for the PM phase of NiO (with the lattice constant $a=4.233$ 
\AA), as $\nu \equiv V/V_0$ \cite{Leonov_2016,Leonov_2020}. In our calculations we assume that the $U$ and $J$ values remain constant upon variation of the lattice volume. 

We use the Curie-Bloch equation $M(T)=M(0)[1-(T/T_N)^\alpha]^\beta$ to fit our results for the temperature dependent on-site magnetization obtained within the spin-polarized DFT+DMFT calculations 
\cite{Kuzmin_2005,Evans_2015}. It allows us to determine the saturated on-site magnetic moment $M(T=0)$, the N\'eel temperature $T_N$, and the parameters $\alpha$ and $\beta$ of the Curie-Bloch 
equation fit (critical exponents). 
Using DFT+DMFT we extract the static uniform spin susceptibility of the Ni $3d$ states in the PM state above $T_N$ as ${\chi_{\mathrm{AFM}}(T) \equiv dM(T)/dH}$, where $M(T)$ is the ordered magnetic 
moment induced by applying a small external magnetic field, $H$. In our calculations, we use three different external magnetic fields, corresponding to splitting of the single-electron energies by 
0.01, 0.005, and 0.002 eV. These values are considered to be small enough to guarantee a linear response regime and to neglect the redistribution of charge density in DFT. For each of theses 
magnetic fields we perform an independent spin-polarized DFT+DMFT calculation. Our result for $\chi_{\mathrm{AFM}}$ is evaluated as an average over these three calculations.
Moreover, we evaluate the Heisenberg exchange couplings of AFM NiO within the spin-polarized DFT+DMFT using the magnetic force theorem 
\cite{Liechtenstein_1987,Katsnelson_2000,Kvashnin_2015,Korotin_2015}. We compute the pressure dependence of the N\'eel temperature and quantify the pressure-temperature phase diagram of NiO.

\section{Results and discussion}

\begin{figure}[tbp!]
\centerline{\includegraphics[width=0.5\textwidth,clip=true]{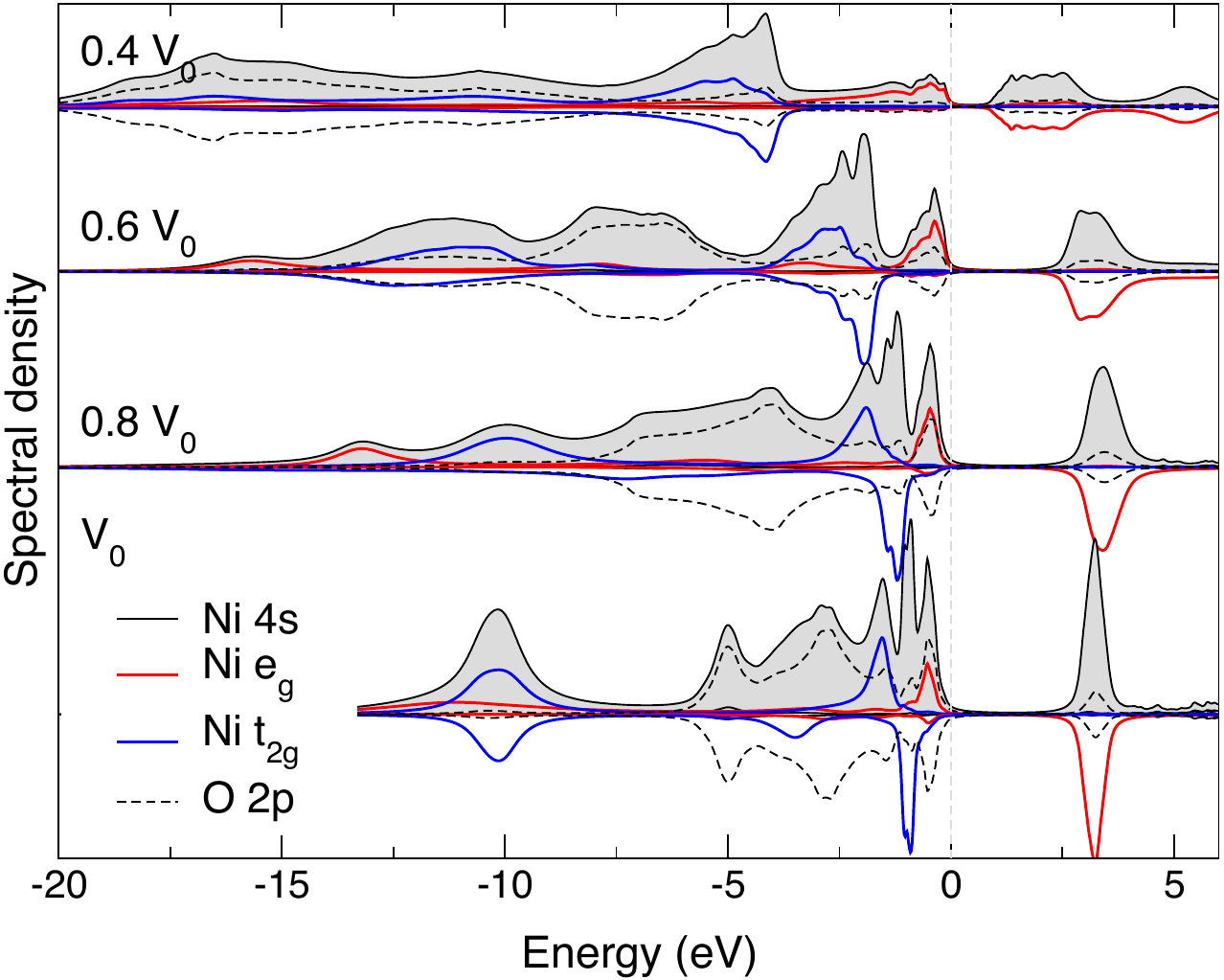}}
\caption{
Our results for the spin-polarized Ni $3d$ and O $2p$ spectral functions obtained by DFT+DMFT with $U=10$ eV and $J=1$ eV for AFM NiO for different lattice compression at a temperature $T=290$ K.
}
\label{Fig_1}
\end{figure}

\begin{figure}[tbp!]
\centerline{\includegraphics[width=0.5\textwidth,clip=true]{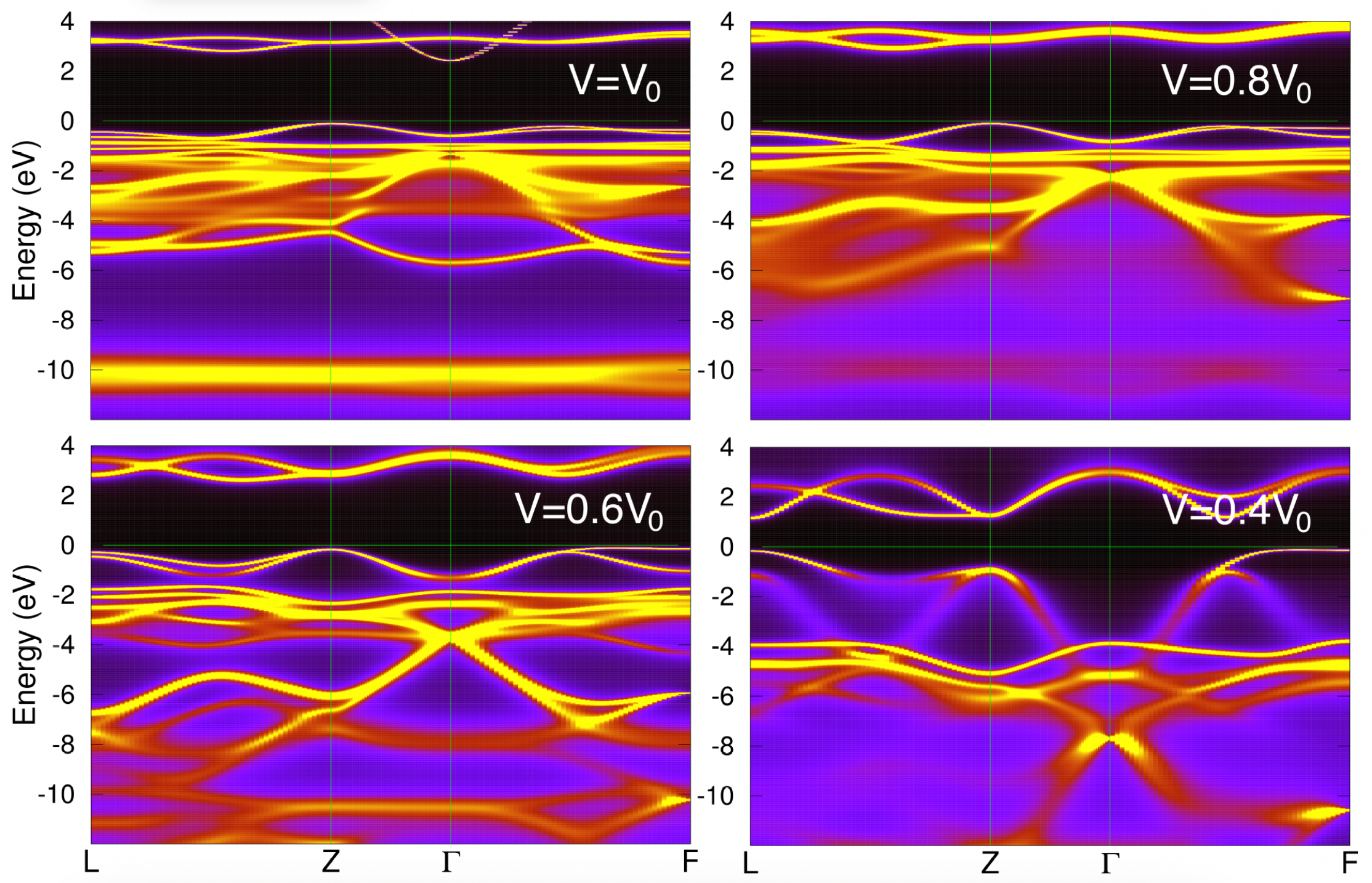}}
\caption{
Our results for the {\bf k}-resolved total spectral function obtained by DFT+DMFT for AFM NiO for different lattice compression at $T=290$ K.
}
\label{Fig_2}
\end{figure}

Using the spin-polarized DFT+DMFT approach we compute the electronic structure and magnetic properties of the cubic $B1$ phase of NiO (for the Hubbard $U=10$ eV and Hund's exchange $J=1$ eV). Our 
results for the orbitally-resolved Ni $3d$ and O $2p$ spectral functions obtained by DFT+DMFT at $T=290$ K as a function of compression are summarized in Fig.~\ref{Fig_1}. In Fig.~\ref{Fig_2} we 
display our results for the corresponding {\bf k}-resolved total spectral functions of AFM NiO. Our results agree well with the previous DFT+DMFT calculations of NiO in the PM state, implying the 
crucial importance of correlations effects of the Ni $3d$ electrons to explain the insulating state of NiO. 

In agreement with experiment, our DFT+DMFT calculations give an insulating solution with a long-range ordering of the Ni $3d$ moments. We establish that upon moderate compression the long-range 
magnetism has no significant effects on the valence band photoemission spectra of NiO \cite{Sawatzky_1984,Shen_1991}. 
Under ambient conditions, we obtain a charge transfer insulator with a large $d$-$d$ energy gap of about 3.3~eV. The top of the valence band is formed by strongly hybridized states originating from 
the majority spin Ni $e_g$ and the O $2p$ orbitals. An indirect gap opens between the top of the valence band at the Brillouin zone (BZ) $Z$ point and the empty parabolic-like Ni $4s$ states located 
near the BZ $\Gamma$ point. The O $2p$ states show a dominant contribution near the Fermi level, in agreement with a charge transfer character of the insulating state 
\cite{Zaanen_1985,Shukla_2003,Schuler_2005}.

The Ni $t_{2g}$ states are fully occupied. The Ni $t_{2g}$ spectral functions show an asymmetric behavior of the majority and minority spin states with a peak located at about -1.6 eV for the 
majority and at -0.9 eV for the minority spin states, respectively. The Ni $3d$ orbitals exhibit a broad structure with strongly incoherent spectral weights near about -10 eV below the Fermi level, 
associated with the lower Hubbard subband.

\begin{figure}[tbp!]
\centerline{\includegraphics[width=0.5\textwidth,clip=true]{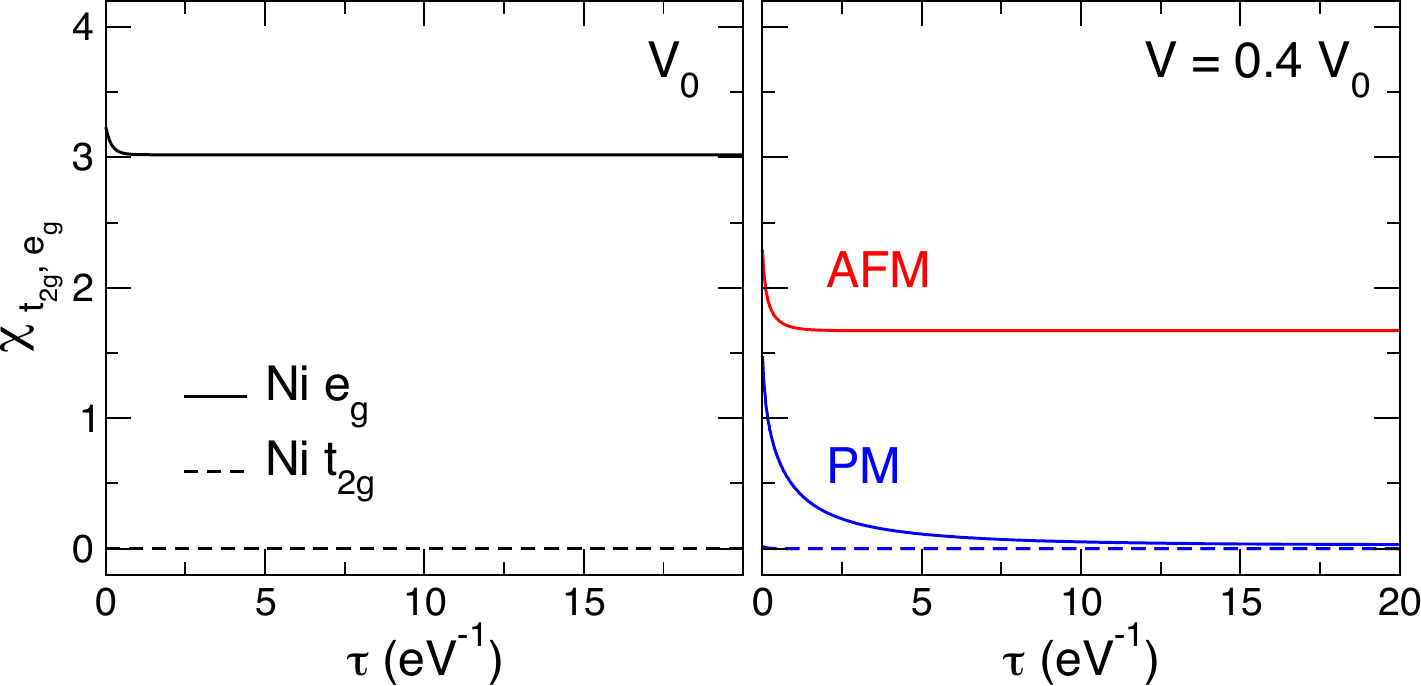}}
\caption{
Orbitally-resolved local spin-spin correlation function $\chi(\tau) \equiv \langle \hat{m}_z(\tau)\hat{m}_z(0)\rangle$ evaluated within DFT+DMFT at $T=290$ K.
}
\label{Fig_3}
\end{figure}

We found that the Ni $e_g$ orbitals are nearly fully spin-polarized and show a large exchange splitting of about 3.6~eV between the (fully occupied) majority and (empty) minority spin Ni $e_g$ 
states. 
The spin-polarized DFT+DMFT calculations yield a large ordered magnetic moment (on-site magnetization of the Ni $3d$ orbitals), $\sim$1.74 $\mu_\mathrm{B}$, implying that Ni ions are in the 
high-spin magnetic state ($S=1$). Our result for the instantaneous local magnetic moments evaluated within DMFT, $\sqrt{\langle \hat{m}^2_z\rangle} \sim 1.81$ $\mu_\mathrm{B}$, is close to the 
fluctuating local moment value, $\sim$1.74~$\mu_\mathrm{B}$, obtained as $\mu=[k_\mathrm{B}T\int \chi(\tau)d\tau]^{1/2}$. Here, $\chi(\tau) \equiv \langle \hat{m}_z(\tau)\hat{m}_z(0)\rangle$ is the 
local spin-spin correlation function evaluated within DMFT on the imaginary time domain $\tau$. This result documents strong localization of the Ni $3d$ orbitals in AFM NiO at ambient pressure, 
similar to that in the PM phase \cite{Leonov_2016,Leonov_2020}, implying the importance of electronic correlations (see Fig.~\ref{Fig_3}).

Our results show a large redistribution of the Ni $3d$ and O $2p$ spectral weights under high pressure. In particular, upon compression to $\sim$0.6~$V_0$, we observe a large shift of the occupied 
part of the O $2p$ states deep below the Fermi level, to about -7.5 eV. It is accompanied by a crossover to a Mott-Hubbard insulating state, with a predominant contribution of the majority Ni $e_g$ 
states near the Fermi level (in comparison to the O $2p$ states). Thus, at high compression, the low-energy excitations have a predominant Ni $d$-$d$ character near the Fermi level, in contrast to 
the $p$-$d$ behavior at ambient pressure. The spectral properties are characterized by a direct gap between the (occupied) majority and (empty) minority spin Ni $e_g$ bands appearing near  the BZ 
$Z$ point. The Ni $4s$ states are shifted by more than $\sim$4 eV above the Fermi level. We also notice a large increase of the bandwidth of the Ni $e_g$ states upon high compression to 
$\sim$0.4~$V_0$ (see Fig.~\ref{Fig_2}). It is particularly seen for the occupied majority spin Ni $e_g$ states located just below the Fermi level. In fact, the majority spin Ni $e_g$ bands show a 
pronounce dispersion with a bandwidth of about 6 eV. The bandwidth of the (empty) minority spin Ni $e_g$ states is sufficiently smaller, about 2 eV. It is important to note that the Ni $e_g$ states 
exhibit high coherence of the spectral weights.

In agreement with our results at ambient pressure, the Ni $e_g$ orbitals are strongly spin-polarized, with a large exchange splitting between the fully occupied majority and empty minority spin Ni 
$e_g$ states, of about 2.7~eV (for the lattice volume $\sim$0.4~$V_0$). The Ni $3d$ states ordered magnetic moments are reduced by about 24\% in comparison to that at ambient pressure, to $\sim$1.32 
$\mu_\mathrm{B}$ (at $T=290$ K). The calculated instantaneous magnetic moments are $\sim$1.54 $\mu_\mathrm{B}$. The corresponding fluctuating local magnetic moments are only 1.3 $\mu_\mathrm{B}$. 
This result implies a drastic suppression of localization of the Ni $3d$ states under high pressure conditions as seen in Fig.~\ref{Fig_3}.

Our DFT+DMFT results for NiO with the $B1$ crystal structure suggest that the antiferromagnetic state of NiO persists up to high compression to $\sim$0.4~$V_0$. Most interestingly, the insulating 
state is found to remain stable up to the highest studied in this work compression value of $\sim$0.4~$V_0$  (at least at $T=290$ K). 
This result differs from the DFT+DMFT calculations for the PM phase of NiO (with the same Hubbard $U=10$ eV and Hund's exchange $J=1$ eV) \cite{Leonov_2016,Leonov_2020}. The previous calculations 
show the Mott IMT in PM NiO upon compression above $\sim$0.54~$V_0$. The Mott transition is accompanied by a crossover from localized to itinerant moment behavior of the Ni $3d$ states. 
Thus, our results for the local spin susceptibility $\chi(\tau)$ for PM NiO show strong delocalization of the Ni $3d$ states upon compression to $\sim$0.4~$V_0$, with $\chi(\tau)$ fast decaying to 
zero at $\tau \sim 20$~eV$^{-1}$ for $T=290$ K (see Fig.~\ref{Fig_3}).
The phase transition is accompanied by divergence of the quasiparticle mass renormalization $m^*/m$ for the Ni $e_g$ states, in accordance with the Brinkman-Rice picture of the Mott IMT 
\cite{Brinkman_1970}. 

In DFT+DMFT we observe a monotonic reduction of the calculated energy gap value of AFM NiO upon high compression, with the calculated energy gap of about 2.8 eV at $\sim$0.6~$V_0$ (at $T=298$ K). It 
corresponds to a 15\% decrease of the gap value upon contraction of the lattice volume to $\sim$0.6~$V_0$). Upon further compression, the gap value sharply drops to 1.1 eV at $\sim$0.4~$V_0$. This 
non-monotonic change suggests a possible crossover in the electronic structure of AFM NiO near $\sim$0.5~$V_0$. Note that this value roughly coincides with the Mott IMT in PM NiO, $\sim0.54$~$V_0$.
An analysis suggests that the insulating state of AFM NiO at the lattice volume $\sim$0.4~$V_0$ can be represented as a (correlated-assisted) Slater insulator driven by long-range magnetic ordering.

\begin{figure}[tbp!]
\centerline{\includegraphics[width=0.5\textwidth,clip=true]{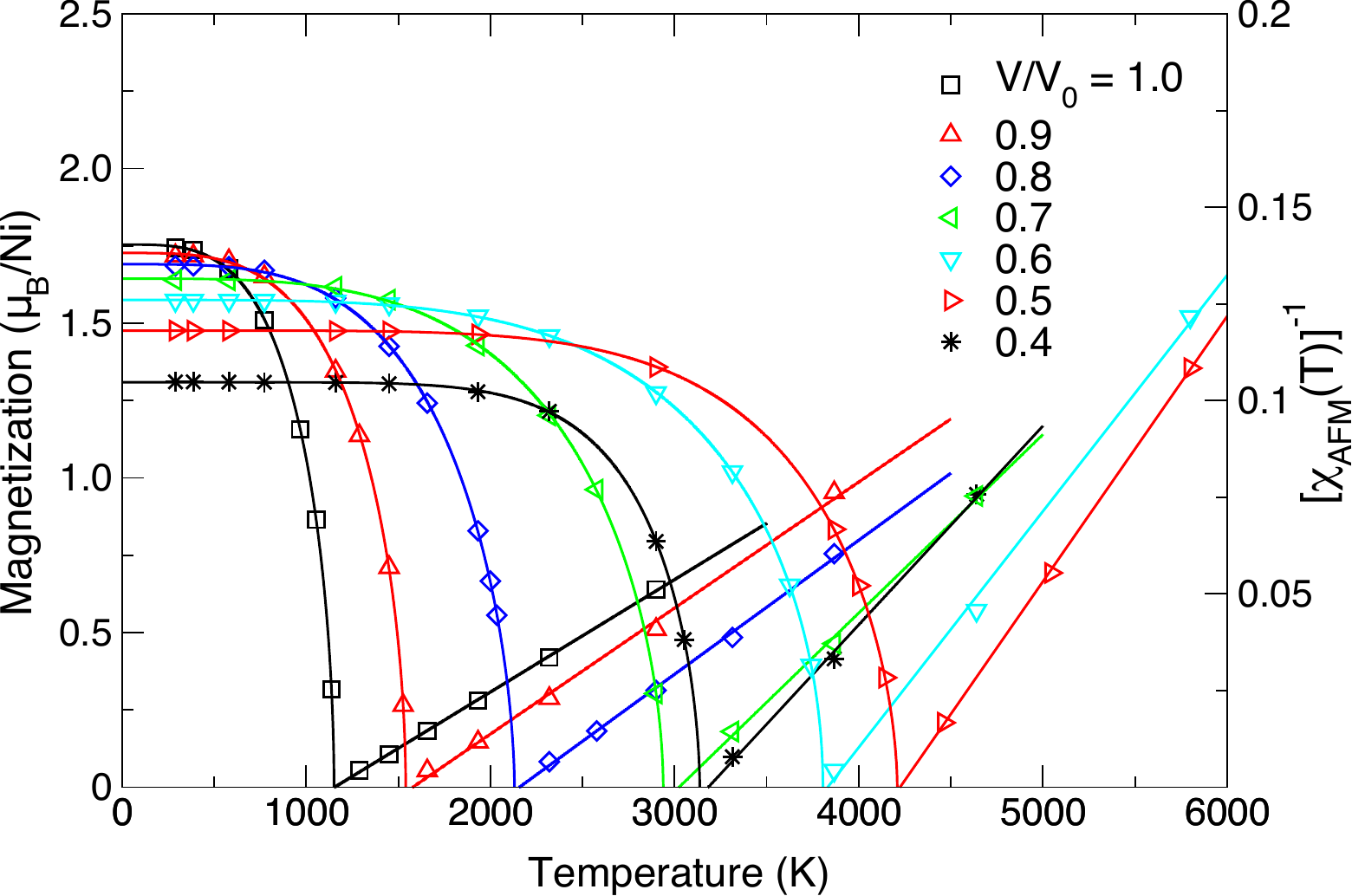}}
\caption{
Our results for the calculated temperature dependence of the ordered magnetic moments (on-site magnetization) and inverse uniform spin susceptibility of the Ni $3d$ states $\chi_\mathrm{AFM}(T)$ 
obtained by DFT+DMFT for NiO as a function of temperature for different compression.
}
\label{Fig_4}
\end{figure}

\begin{figure}[tbp!]
\centerline{\includegraphics[width=0.5\textwidth,clip=true]{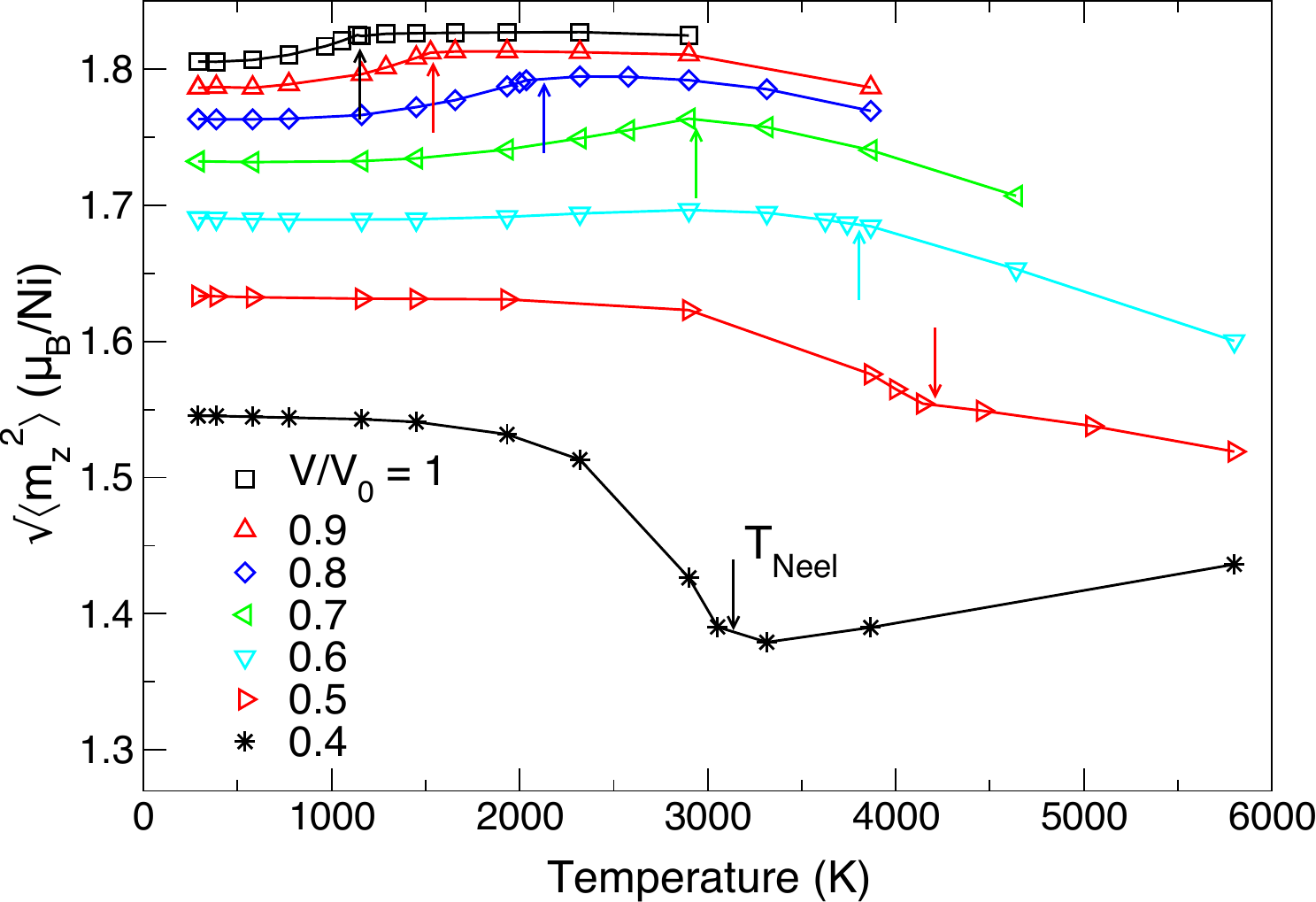}}
\caption{
The Ni $3d$ instantaneous local magnetic moments for different lattice compression ($V/V_0$) and temperature obtained using DFT+DMFT. The calculated N\'eel temperatures at given compression are 
depicted by arrows.
}
\label{Fig_5}
\end{figure}

Using the DFT+DMFT method it becomes possible to address the finite-temperature aspects of magnetic behavior of strongly correlated materials. We use DFT+DMFT to compute the temperature dependence 
of ordered  and local magnetic moments of NiO under pressure. We fit our results for the temperature dependent ordered magnetic moments to the Curie-Bloch equation 
$M(T)=M(0)[1-(T/T_N)^\alpha]^\beta$ and determine the corresponding model parameters: the saturated on-site magnetization $M(T=0)$,  N\'eel temperature $T_N$, and critical exponents $\alpha$ and 
$\beta$. In addition, using DFT+DMFT we extract the static uniform spin susceptibility of the Ni $3d$ states in the PM state above $T_N$ as ${\chi_{\mathrm{AFM}}=dM/dH}$. 
As a result, we evaluate the pressure dependence of the N\'eel temperature of NiO. Our results are summarized in Fig.~\ref{Fig_4}. The temperature dependence of the local magnetic moments for 
different lattice compression are shown in Fig.~\ref{Fig_5}. 

Our results for the inverse uniform magnetic susceptibility of the Ni $3d$ states, $\chi_{\mathrm{AFM}}^{-1}$, show the Curie-Weiss behavior at high temperature (see Fig.~\ref{Fig_4}). Upon heating, 
we observe a phase transition to the PM state, associated with a suppression of the long-range AFM state (the ordered magnetic moments turn to zero). On the other hand, the local magnetic moments 
remain finite above $T_N$ in the PM states and show a nontrivial behavior near $T_N$. Our DFT+DMFT results at ambient pressure show a rather large $T_N$ value of $\sim$1150 K, i.e, the calculated 
$T_N$ is about twice overvalued in comparison to the experimental value of $\sim$523~K. This implies the crucial importance of the non-local correlations, e.g., the effects associated with long 
wavelength spin waves, missing in the single-site DFT+DMFT approach \cite{Held_2011,Tomczak_2012,Boehnke_2016,Choi_2016,Nilsson_2017,Tomczak_2017,Zgid_2017,
Iskakov_2020,Zhu_2021,Toschi_2007a,Toschi_2007b,Katanin_2019,
Rubtsov_2008,Rubtsov_2009,Rohringer_2018,Cunningham_2023}.

We observe a monotonic decrease of the saturated on-site magnetization $M(T=0)$ upon compression. This result is in qualitative agreement with a compressional behavior of the instantaneous local 
magnetic moments that exhibit a nontrivial behavior with temperature, with a clear crossover near to the calculated $T_N$ (see Fig.~\ref{Fig_5}). It is important to distinguish two distinct regimes 
under pressure. In particular, for $V \geq 0.6~V_0$ we observe a monotonous \emph{increase} of the local magnetic moments upon heating to about the N\'eel temperature, followed by decreasing of the 
calculated local magnetic moments upon heating far above $T_N$ (see Fig.~\ref{Fig_5}). In contrast to this, under high compression, for $V<0.6~V_0$, the local magnetic moments monotonically 
\emph{decrease} upon heating to about $T_N$. For $ \sim$0.4~$V_0$, the local magnetic moments are seen to increase above $T_N$, in accordance with the spin-fluctuation theory for itinerant magnetism 
\cite{Mori_1978,Hasegawa_1983}. This result corroborates with the above proposed crossover to the correlated-assisted Slater insulating state (band insulator driven by long-range magnetic ordering) 
under high compression, for $V<0.5~V_0$.

\begin{figure}[tbp!]
\centerline{\includegraphics[width=0.46\textwidth,clip=true]{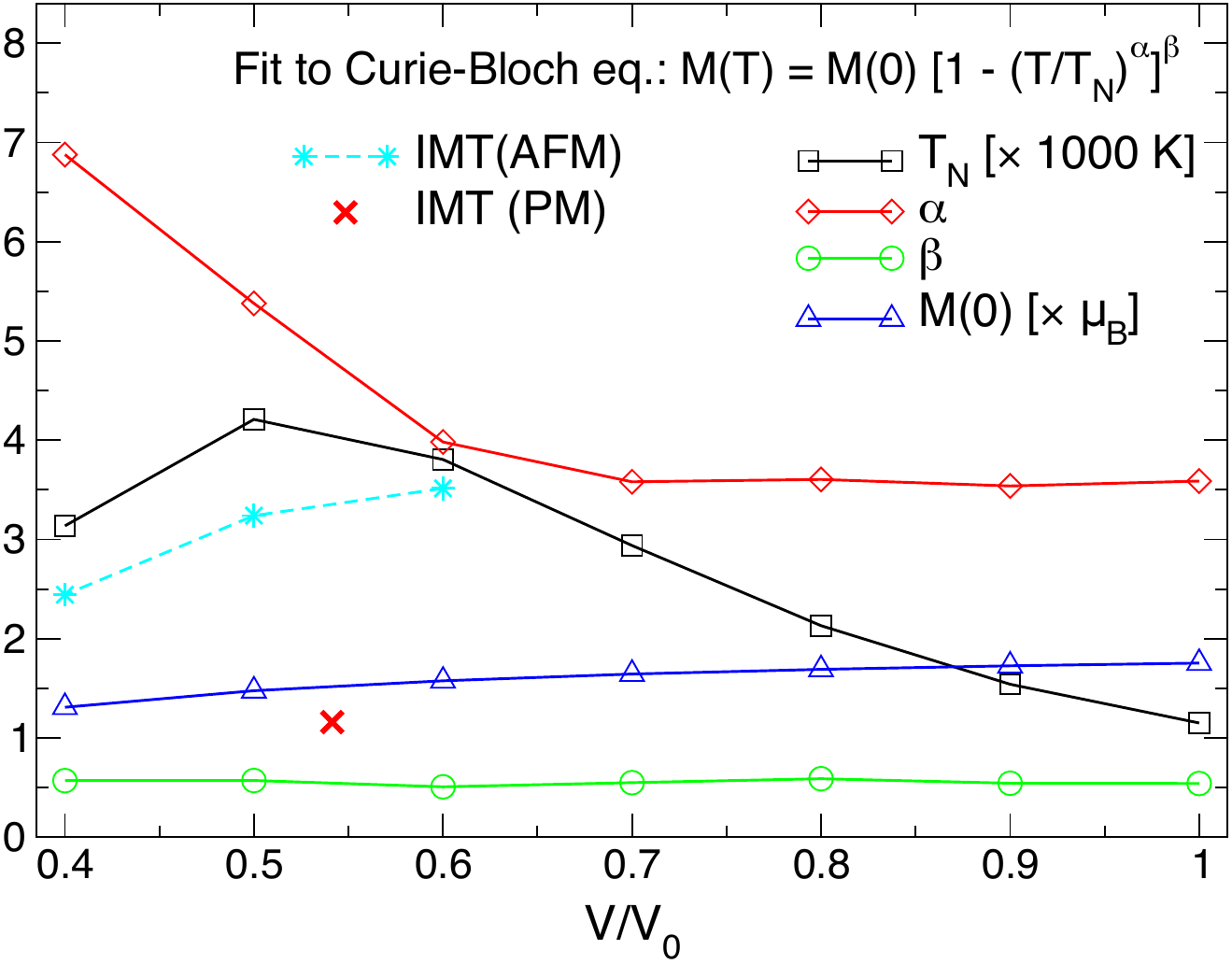}}
\caption{
Our results of the fitting of the calculated temperature dependence of the ordered magnetic moments to the Curie-Bloch equation for different compression of the lattice $V$/$V_0$. The calculated 
N\'eel temperature $T_N$, parameters $\alpha$ and $\beta$, and saturated ordered magnetic moment $M(T=0)$ are shown. The IMT evaluated within the spin-polarized DFT+DMFT calculations for AFM NiO is 
depicted as IMT(AFM). The previously obtained DFT+DMFT result for PM NiO (with imposed PM symmetry) is marked as IMT (PM) \cite{Leonov_2016,Leonov_2020}.
}
\label{Fig_6}
\end{figure}

Our result for the calculated N\'eel temperature shows a non-monotonic behavior of $T_N$ under pressure. In fact, the calculated $T_N$ has a maximum located near $\sim$0.5~$V_0$ at which $T_N$ is by 
about four times larger than that at ambient compression.
Note that this behavior qualitatively agrees with the pressure-temperature phase diagram of the half-filled single-band Hubbard model and can be interpreted as a crossover from a localized to 
itinerant moment behavior in NiO under pressure \cite{Rohringer_2011,Kunes_2011,LeBlanc_2015,Schafer_2015,Ayral_2016,Rohringer_2016,Harkov_2021,
Iskakov_2022,Stobbe_2022,Lenihan_2022}. Indeed, AFM NiO under pressure exhibit a crossover between the strong to weak coupling regimes,
associated with a transition from the Mott- to Slater-type (band) insulator. It is particularly remarkable that this crossover correlates with a qualitative change in the behavior of the parameter 
$\alpha$ in the Curie-Bloch equation for the lattice volumes below $\sim$0.6~$V_0$ (see Fig.~\ref{Fig_6}). While being nearly constant upon small compression, $\alpha$ drastically increases by about 
two times upon high compression for the lattice volumes below $\sim$0.6~$V_0$. At the same time, the critical exponent $\beta$ remains nearly constant, about 0.55, in reasonable agreement with the 
mean-field value of 0.5. 

Our analysis also suggests that upon high compression below $\sim$0.6~$V_0$ AFM NiO undergoes an insulator-to-metal phase transition with the transition temperature sufficiently below $T_N$. This 
suggests an appearance of the antiferromagnetic \emph{metallic} phase of NiO at the onset of magnetic ordering, at high compression below $\sim$0.6~$V_0$. It is particularly interesting that under 
high compression of $\sim$0.4~$V_0$ the IMT temperature is more than by twice larger than the N\'eel temperature of NiO obtained at ambient pressure. Thus, the long-range magnetically ordered 
insulating state of NiO is robust and persists up to high compression.

\begin{figure}[tbp!]
\centerline{\includegraphics[width=0.5\textwidth,clip=true]{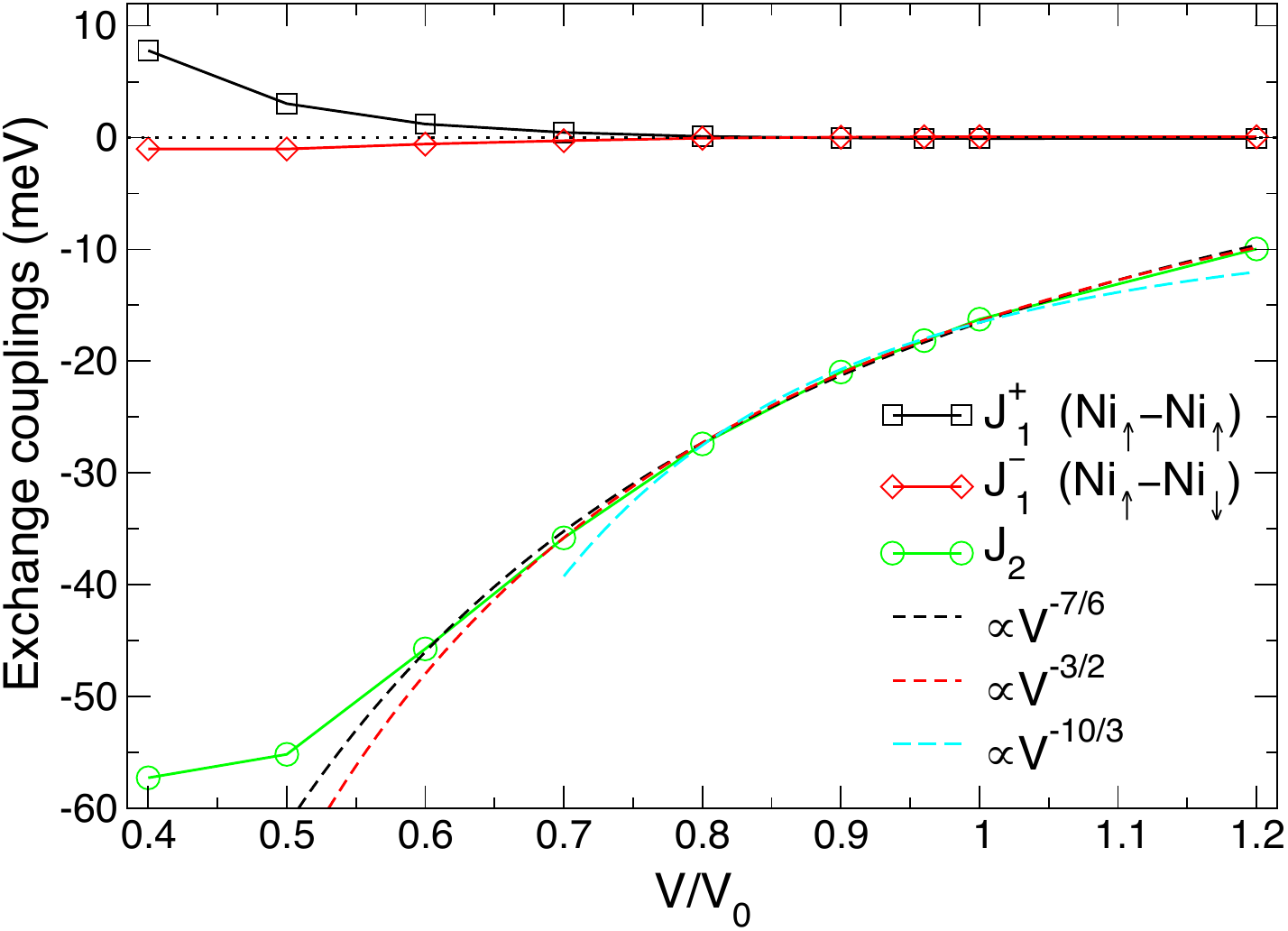}}
\caption{
Our results for the nearest neighbor ($J_1^+$ and $J_1^-$) and next-nearest neighbor ($J_2$) exchange couplings of NiO calculated using the magnetic force theorem within the spin-polarized DFT+DMFT 
for different lattice compression.
}
\label{Fig_7}
\end{figure}

Next, we calculate the Heisenberg exchange couplings of AFM NiO using the magnetic force theorem within the spin-polarized DFT+DMFT calculations at $T=290$ K. Our results are displayed in 
Fig.~\ref{Fig_7}. 
In agreement with the Goodenough-Kanamori-Anderson rules \cite{Goodenough_1963,Kanamori_1959,Anderson_1950} the predominant interaction is a large antiferromagnetic exchange between the next-nearest 
neighbor Ni ions, $J_2$. 
It is associated with a 180$^\circ$ Ni-O-Ni superexchange, driven by the overlap of the Ni$^{2+}$ $e_g$ and O $2p$ orbitals.
The interactions between the nearest neighbor Ni ions ($J_1$) are much smaller and presumably ferromagnetic, caused by the $90^\circ$ Ni-O-Ni superexchange interactions.
Note that there are two distinct values of $J_1$ denoted by $J_1^+$ and $J_1^-$, where the plus and the minus signs denoting coupling between neighbors having parallel and antiparallel spin 
orientations, respectively. 
For the ambient pressure (for the lattice constant $a=4.233$ \AA\ obtained within DFT+DMFT) our calculations yield $J_2 \sim -16.24$~meV, in agreement with experimental estimates from inelastic 
neutron scattering, -19.01 meV (at  $T = 78$~K), and from the low-temperature powder susceptibility measurements, -17.32~K. For the experiment lattice constant (for the cubic $Fm\bar{3}m$ crystal 
structure with $a=4.17$ \AA) our calculations give $J_2 \sim 18.16$~meV, implying a sensitive dependence of $J_2$ upon the lattice volume. Our results for the 
$J_1^+$ and $J_1^-$ are 0.08 meV and -0.08 meV, respectively (at ambient pressure).
It is also interesting to note that the mean-field estimate of the critical temperature for the type-II antiferromagnetic ordering as $k_BT_N=6J_2 \frac{1}{3}S(S+1)$ gives a large value of 754~K, 
which is comparable with the N\'eel temperature value obtained by DFT+DMFT, $T_N\sim 1150$~K.
Moreover, this suggests that magnetic correlations in NiO are strong and can persists as short range magnetic correlations far above $T_N$. This result is in agreement with the diffuse magnetic 
scattering study of NiO which shows the presence of strong spin correlations above the ordering temperature, even at $T \simeq 1.5~T_N$ \cite{Hutchings_1972}.

It is expected that pressure leads to an increase of the superexchange coupling, derived according to second-order perturbation theory as $|J| \propto t^2/U$. In fact, our results show that the 
next-nearest neighbor exchange coupling monotonically increase (by module) upon lattice compression, by more than a factor of two upon compression to $\sim$0.5~$V_0$. As a result, this leads to the 
enhancement of $T_N$ upon compression to its maxima at about $\sim$0.5~$V_0$. Most interestingly, upon further compression above $\sim$0.5~$V_0$, $T_N$ is found to decrease significantly, while 
$J_2$ monotonically increases to about -57.25 meV at $\sim$0.4~$V_0$. We connect this unexpected behavior (a large decrease of $T_N$ upon increasing of the unfrustrated exchange coupling $J_2$) with 
a crossover to weak coupling regime (with itinerant magnetic moments). We also note a drastic increase of the FM nearest neighbor coupling $J_1^+$, while the AFM $J_1^-$ exchange remains nearly 
constant. 
Moreover, our fit of the pressure dependence of the exchange parameter gives a rather good description for $J \propto V^{-\gamma}$ in a broad range of compressions with $\gamma$ ranging from 7/6 to 
3/2, in contrast to the Bloch's empirical $J \propto V^{-10/3}$ low \cite{Bloch_1966,Massey_1990}. 
In fact, the latter is valid only in a relatively narrow range of $\gamma$ between $\sim$0.75~$V_0$ to 1.05~$V_0$.
It is remarkable that our empirical fit is applied for a broad range of compressions between $\sim$0.6~$V_0$ to 1.2~$V_0$, breaks upon high compression below $\sim 0.6$~$V_0$. 

Our results for the electronic structure and magnetic properties of NiO are in agreement with experiments, implying that the properties of AFM NiO to a large extent can be explained by ignoring the 
rhombohedral crystal distortions. This suggests the crucial importance of the effects of strong correlations over the band-structure details. The minor rhombohedral distortion of the cubic structure 
seems to be important to explain magnetic anisotropy of NiO. Its theoretical understanding represents a challenging topic for the future.

\section{Summary and conclusions}

In conclusion, using the DFT+DMFT method we have calculated the pressure-induced evolution of the electronic structure, magnetic state, and exchange couplings of the prototypical correlated 
insulator, NiO. This allows us to determine the pressure-temperature phase diagram of NiO. 
In agreement with experiment, our DFT+DMFT results establish that the long-range magnetism has no significant effects on the valence band photoemission spectra of NiO upon moderate compressions. 
This implies the importance of correlations effects of the Ni $3d$ electrons which explain the insulating state of NiO.
Our results document a pressure-driven crossover of the electronic sates of AFM NiO from a charge transfer (under ambient pressure) to the Mott-Hubbard insulating character of the band gap. It is 
accompanied by a graduate monotonic reduction of the calculated energy gap value of AFM NiO, by about 15\% upon compression to $\sim$0.6~$V_0$.
Upon further compression, the gap value sharply drops to 1.1 eV. This non-monotonic change of the electronic structure of AFM NiO near $\sim$0.5~$V_0$ is accompanied by a large redistribution of the 
Ni $3d$ and O $2p$ spectral weights. Moreover, we found that at such compression the PM phase of NiO is metallic. In fact, our DFT+DMFT calculations of the PM phase of NiO upon compression above 
$\sim$0.54~$V_0$ give the Mott IMT which is accompanied by delocalization of the Ni $3d$ electrons.
In contrast to this, the antiferromagnetic insulating state is found to remain stable up to the highest considered in this study compression value of $\sim$0.4~$V_0$ (assuming the cubic $B1$ crystal 
structure of NiO). 
Our results therefore suggest that NiO makes a pressure-driven crossover from the Mott localized to itinerant band insulator behavior, while the insulating state of AFM NiO at the lattice volume 
$\sim$0.4~$V_0$ can be represented as a correlated-assisted Slater insulator driven by long-range magnetic ordering.

The Mott to band insulator crossover in AFM NiO is accompanied by a non-monotonic behavior of the calculated $T_N$ upon compression. In qualitative agreement with the temperature vs $U/t$ phase 
diagram of the half-filled single-band Hubbard model, the maximum of $T_N$ under pressure correlates with the Mott IMT in the PM state of NiO. We associate this behavior with a crossover from strong 
to weak coupling regime.
While the long-range magnetic ordering and insulating state of NiO is robust and persists up to high compressions to $\sim$0.4~$V_0$, our analysis propose the appearance of the antiferromagnetic 
metallic phase of NiO at the onset of the magnetic ordering, at high compression below $\sim$0.6~$V_0$.
Our results show the crucial importance of the non-local correlation effects (and, in particular, the presence of long wavelength spin waves) to explain the magnetic properties of NiO. 
In order to explain the possible pressure-driven metallization of NiO one should take into account structural transformations of NiO upon high compression. 
This challenging topics deserves further theoretical and experimental analysis.

\begin{acknowledgments}
The DFT+DMFT calculations were supported by the Russian Science Foundation (Project No. 19-72-30043). Our theoretical analysis of the electronic structure and magnetic properties was supported by 
the Ministry of Science and Higher Education of the Russian Federation, project No. 122021000038-7 (theme ``Quantum'').
\end{acknowledgments}

\end{document}